\documentclass[prl,twocolumn,showpacs,floatfix,superscriptaddress]{revtex4-1}
\usepackage{graphicx}
\usepackage{amsmath}
\usepackage{amsfonts}
\usepackage{amssymb}
\usepackage{hyperref}
\hypersetup{bookmarks,colorlinks,citecolor=red,filecolor=blue,urlcolor=blue}
\usepackage{subfigure}

\begin{document}

\title{A Passive Phase Noise Cancellation Element}%
\author{Eyal Kenig}
\email[Corresponding author:\ ]{eyalk@caltech.edu}
\affiliation{Kavli Nanoscience Institute and Condensed Matter Physics, California Institute of Technology, MC 149-33, Pasadena, California 91125, USA}
\author{M. C. Cross}
\affiliation{Kavli Nanoscience Institute and Condensed Matter Physics, California Institute of Technology, MC 149-33, Pasadena, California 91125, USA}
\author{Ron Lifshitz}
\affiliation{Raymond and Beverly Sackler School of Physics and Astronomy, Tel Aviv University, 69978 Tel Aviv, Israel}
\author{R. B. Karabalin}
\affiliation{Kavli Nanoscience Institute and Condensed Matter Physics, California Institute of Technology, MC 149-33, Pasadena, California 91125, USA}
\author{L. G. Villanueva}
\affiliation{Kavli Nanoscience Institute and Condensed Matter Physics, California Institute of Technology, MC 149-33, Pasadena, California 91125, USA}
\author{M. H. Matheny}
\affiliation{Kavli Nanoscience Institute and Condensed Matter Physics, California Institute of Technology, MC 149-33, Pasadena, California 91125, USA}
\author{M. L. Roukes}
\affiliation{Kavli Nanoscience Institute and Condensed Matter Physics, California Institute of Technology, MC 149-33, Pasadena, California 91125, USA}
\begin{abstract}
We introduce a new method for reducing phase noise in oscillators, thereby improving their frequency precision. The noise reduction device consists of a pair of coupled nonlinear resonating elements that are driven parametrically by the output of a conventional oscillator at a frequency close to the sum of the linear mode frequencies. Above the threshold for parametric response, the coupled resonators exhibit self-oscillation at an inherent frequency. We find operating points of the device for which this periodic signal is immune to frequency noise in the driving oscillator, providing a way to clean its phase noise. We present results for the effect of thermal noise to advance a broader understanding of the overall noise sensitivity and the fundamental operating limits.
\end{abstract}

\date{\today}


\maketitle

The emergence of self-oscillation has a major scientific significance as a widespread phenomenon in physics, chemistry, and biology \cite{strogatz}. Oscillators are also extremely useful, frequently appearing as crucial elements in the electrical devices that surround us in our highly technological environment. Essentially, oscillators are devices generating a periodic signal at an inherent frequency, whose primary function is therefore to provide a time or a frequency reference.  An ideal self-sustained oscillator is mathematically described as a limit-cycle in the phase space of dynamical variables, or equivalently as a periodic solution of a set of autonomous differential equations, independent of an external time reference. The ideal oscillator can thus be described in terms of a steadily increasing phase variable corresponding to the phase space point advancing around the limit-cycle, with a $2\pi$ phase change corresponding to a period of the motion. This phase is highly sensitive to additional stochastic terms, or noise, in the equations of motion, as the appearance of periodicity without an external time reference implies the freedom to drift along the phase direction. The stochastic phase dynamics lead to a broadening of the peaks in the power spectrum of the oscillator output, which are perfectly discrete in the ideal case \cite{lax67,Demir00}, and to a degradation of the performance as a time or frequency reference \cite{vig}. Thus, an essential task in the design of a good oscillator is to reduce the effects of the noise, present in the system, on the oscillator phase.

\begin{figure}[]
\begin{center}
   \includegraphics[width=1\columnwidth]{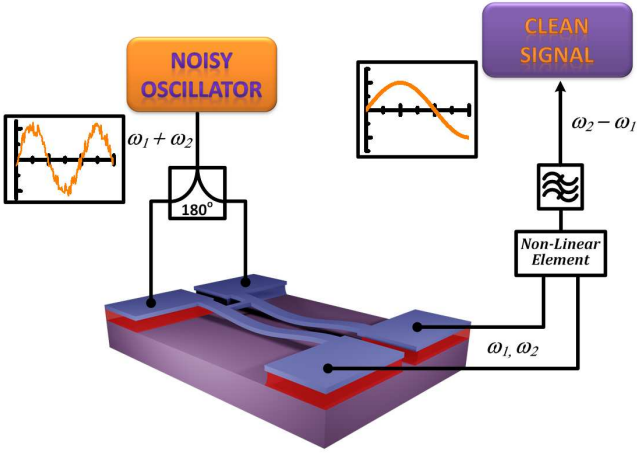}
   \end{center}
   \caption{\label{fig1} (Color online) An illustration of the phase noise cancellation scheme. An oscillator produces a signal with a noisy frequency around $\omega_1+\omega_2$. This signal parametrically drives a pair of coupled NEMS or MEMS beams with a relative phase of $180^\circ$. The output signal at a frequency around $\omega_2-\omega_1$ is given by squaring and filtering.}
\end{figure}

In this letter we propose a general scheme to reduce, or even eliminate, the noise in the output of an oscillator by passing the signal through a second passive noise cancellation device, rather than manipulating the oscillator itself. It is therefore broadly applicable to enhance the performance of existing oscillator designs. It is also valuable from a basic physics perspective, eliminating the need to analyze the $\emph{active}$ resonator-amplifier feedback system: instead, the noise performance is mapped to a $\emph{passive}$ element, whose fundamental stochastic properties are more amenable to the powerful tools of statistical physics \cite{clerk}. The context of our work is to use nanoscale or microscale resonators to build high precision oscillators as illustrated in Fig.~\ref{fig1}, although our scheme applies more generally. The nano- or micro-electromechanical systems (NEMS or MEMS) implementation shown in Fig.~\ref{fig1} consists of a pair of coupled resonating elements that are parametrically-driven with a noisy frequency near the sum of their linear mode frequencies. As explained below, it produces a signal with reduced noise at a frequency near the difference of these linear frequencies. The generation of a two-frequency signal by parametric excitation at the frequency sum is called non-degenerate parametric excitation, a phenomenon that has been studied in mechanical systems \cite{Turner03}, but is more common in the context of optical parametric oscillators \cite{kroll62,*wong90}, where our scheme also applies.

The ability of an oscillator based on a nonlinear resonator to suppress phase noise was demonstrated by Yurke $\textit{et al.}$ \cite{yurkePra}, who studied a nonlinear beam resonator driven into self-oscillation through an active feedback loop, composed of an amplifier driven to saturation and a phase shifter. They showed that if the resonator is operated at the critical Duffing point, by applying the minimal drive required for making the amplitude \textit{vs.}~frequency curve multivalued, and by choosing the phase shift in the feedback loop between the signal and the drive to be $2\pi/3$, the noise in the phase of the fedback signal has no effect on the oscillator phase. This was understood in terms of the insensitivity of the resonator frequency to the drive phase at this specific operating point. Here we generalize this understanding of noise reduction in the feedback oscillator, and show that in the present case the output phase noise due to phase noise in the drive may be eliminated by operating at any points for which the output frequency itself is insensitive to the drive frequency.

We initiate our analysis by modeling a pair of coupled resonators, like the ones shown in Fig.~\ref{fig1}, by two dimensionless equations of motion
\begin{eqnarray}\label{eom}
\ddot{x}_{n} &+& x_{n}+x_{n}^{3}+\epsilon\dot{x}_{n}+\eta x_{n}^{2}\dot{x}_{n}+\epsilon h_{n}x_{n}\cos(\omega_{p}t)\nonumber\\&+&\Delta(x_{n}-x_{k}) = 0,
\end{eqnarray}
with $(n,k)=(1,2)$ and $(2,1)$. For details on the use of such equations for modeling NEMS and MEMS devices see Lifshitz and Cross~\cite{LC03,LCreview}. We only wish to highlight the following points. The resonators are taken to be identical, having a resonant frequency $\omega_0$, which has been scaled out, although considering different resonant frequencies would not qualitatively change our results. The nonlinearity of the resonators originates from both the elastic restoring force and damping mechanism. In micro- and nano-scale resonators the linear damping is typically weak, $\epsilon\ll1$, where $1/\epsilon$ is the quality factor of the resonator, and correspondingly a small drive amplitude is sufficient to excite them. Accordingly, each resonator is parametrically excited with a drive amplitude $\epsilon h_n$. In this regime, we can focus on the slow-time modulation of the basic oscillatory motion of the in-phase and out-of-phase modes of the coupled resonators, described by a pair of complex equations for the corresponding mode amplitudes $A_n(T)$, where $T=\epsilon t$ is a slow time scale. In the regime of strong coupling, where the linear mode splitting is much larger than the resonator bandwidth, these equations are
 \begin{subequations}
 \label{amp_eq}
 \begin{eqnarray}
     \frac{dA_{1}}{dT}&=&-\frac{1}{2}(1+i\Omega_p)A_{1}+i\frac{g}{4\omega_1}A_{2}^{*}\nonumber\\
     &+&\frac{(3i-\eta \omega_{1})}{8\omega_1}(|A_{1}|^{2}A_{1}+2|A_{2}|^{2}A_{1}),\\
       \frac{dA_{2}}{dT}&=&-\frac{1}{2}(1+i\Omega_p)A_{2}+i\frac{g}{4\omega_2}A_{1}^{*}\nonumber\\
       &+&\frac{(3i-\eta \omega_{2})}{8\omega_2}(|A_{2}|^{2}A_{2}+2|A_{1}|^{2}A_{2}),
 \end{eqnarray}
 \end{subequations}
where $\omega_{1}^{2}=1$ and $\omega_{2}^{2}=1+2\Delta$ are the linear frequencies of the in-phase and out-of-phase modes, respectively, $\epsilon\Omega_{p}=\omega_{p}-\omega_{1}-\omega_{2}$ is the small difference between the drive frequency and the sum of the linear mode frequencies, and $g=(h_1-h_2)/2$. Similar slow equations for pairs of coupled resonators were recently introduced to study chaotic dynamics \cite{kar09,*Kenig11} and to analyze the so-called Bifurcation-Topology Amplifier~\cite{BTA}.

\begin{figure}[]
\begin{center}
   \subfigure[]{
   \includegraphics[width=0.48\columnwidth]{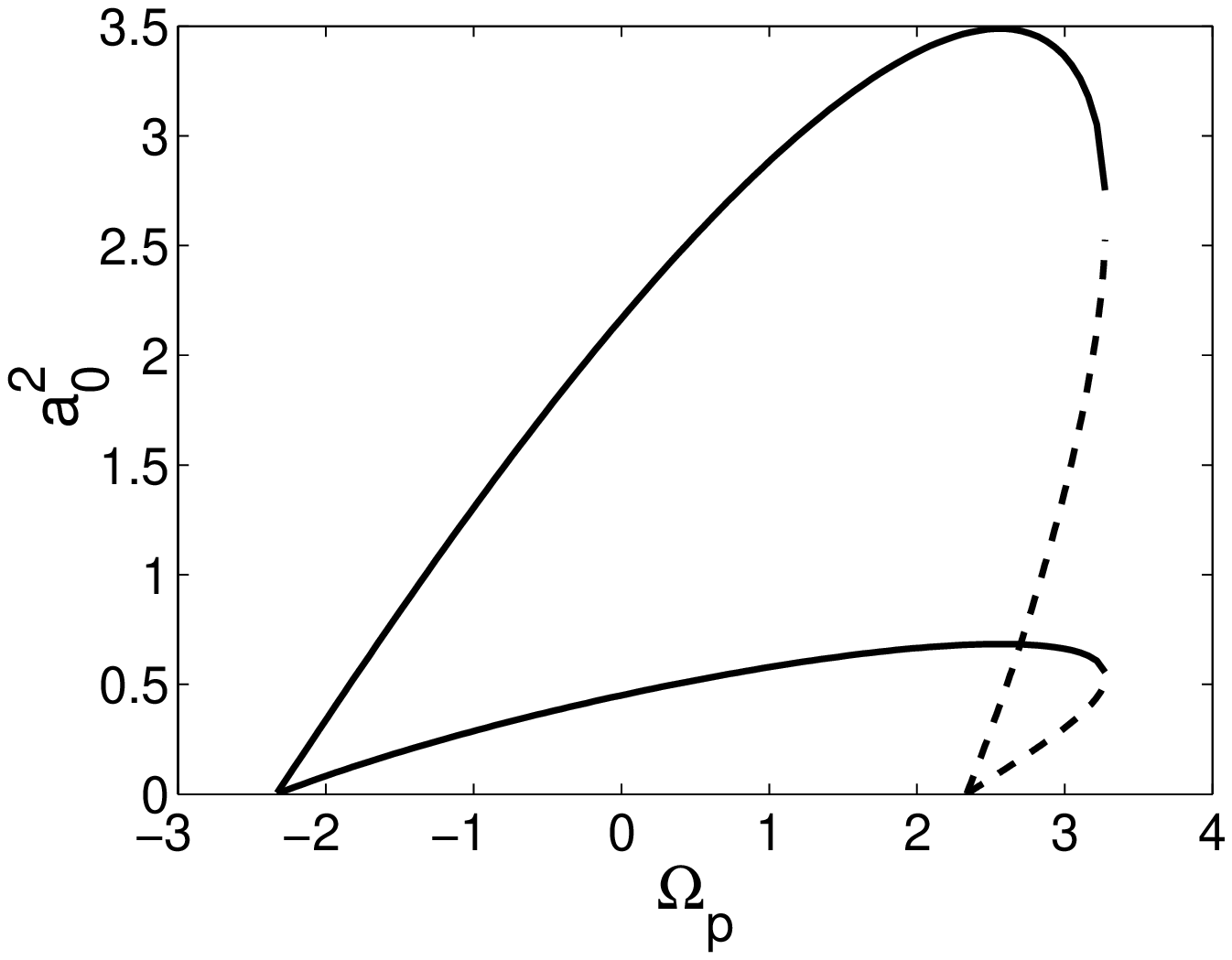}}
   \subfigure[]{
   \includegraphics[width=0.48\columnwidth]{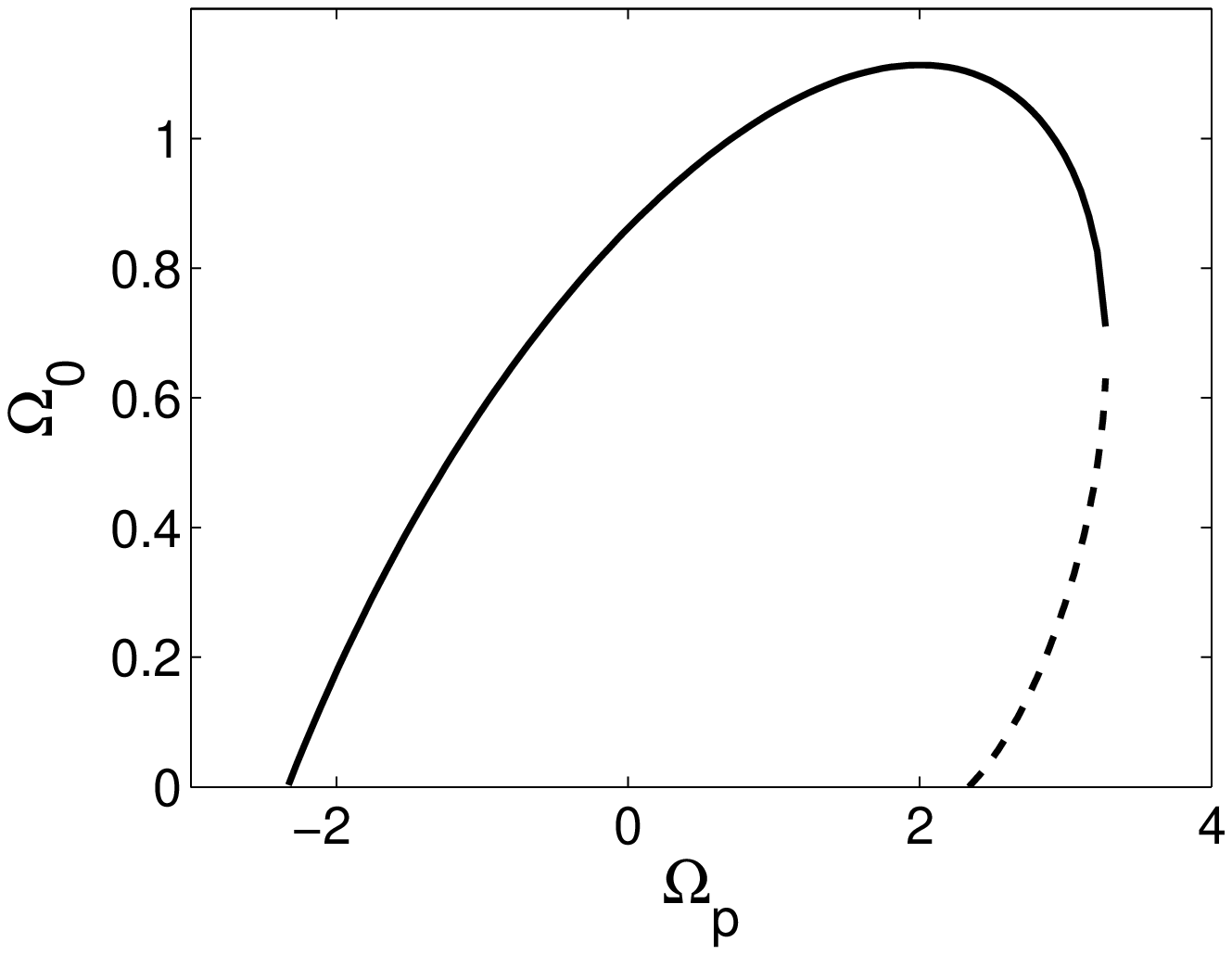}}
   \end{center}
   \caption{\label{fig2} The periodic solutions of Eqs.~(\ref{amp_eq}). (a) The squared mode amplitudes (the in-phase mode has the larger amplitude); (b) Twice the frequency of the periodic solutions. Solid and dashed lines are stable and unstable solutions, respectively. The parameters are $\Delta=7$, $g=10$, and $\eta=1$.}
\end{figure}

In magnitude-phase coordinates,
$A_n=a_ne^{i\phi_n}$, Eqs.~(\ref{amp_eq}) can be transformed into four dynamical equations for the variables  $a_1$, $a_2$, $\Phi=\phi_1+\phi_2$, and $\Psi=\phi_1-\phi_2$,
\begin{subequations}\label{polar}
 \begin{eqnarray}
   \frac{da_{1}}{dT} &=& -\frac{a_{1}}{2}+\frac{ga_{2}}{4\omega_{1}}\sin\Phi-\frac{\eta}{8}(a_{1}^{3}+2a_{2}^{2}a_{1})
   \nonumber\\&=&f_1(a_1,a_2,\Phi)\label{amp1}, \\
   \frac{da_{2}}{dT} &=& -\frac{a_{2}}{2}+\frac{ga_{1}}{4\omega_{2}}\sin\Phi-\frac{\eta}{8}(a_{2}^{3}+2a_{1}^{2}a_{2})\nonumber\\
   &=&f_2(a_1,a_2,\Phi),\label{amp2} \\
   \frac{d\Phi}{dT} &=& -\Omega_{p}+\frac{g}{4}\cos\Phi\left(\frac{a_{2}}{a_{1}\omega_{1}}+\frac{a_{1}}{a_{2}\omega_{2}}\right)\nonumber\\
   &+&\frac{3}{8}\left(\frac{a_{1}^{2}+2a_{2}^{2}}{\omega_{1}}+\frac{a_{2}^{2}+2a_{1}^{2}}{\omega_{2}}\right)= f_3(a_1,a_2,\Phi)\label{phi},\\
   \frac{d\Psi}{dT} &=& \frac{g}{4}\cos\Phi\left(\frac{a_{2}}{a_{1}\omega_{1}}-\frac{a_{1}}{a_{2}\omega_{2}}\right)\nonumber\\
   &+&\frac{3}{8}\left(\frac{a_{1}^{2}+2a_{2}^{2}}{\omega_{1}}-\frac{a_{2}^{2}+2a_{1}^{2}}{\omega_{2}}\right)= f_4(a_1,a_2,\Phi)\label{psi}.
 \end{eqnarray}
\end{subequations}
The right hand sides of these equations depend only on the three variables $a_1$, $a_2$ and $\Phi$. The absence of a dependence on $\Psi$ reflects the fact that Eqs.~(\ref{amp_eq}) are unchanged by the transformation $(\phi_1,\phi_2)\rightarrow(\phi_1+\beta,\phi_2-\beta)$. This property of the non-degenerate response to parametric excitation is well known \cite{graham70,*reid89,*sanders90,*drummond90,*slosser94}, and does not occur in the degenerate case, where $\omega_1=\omega_2$, in which both phases are fixed, corresponding to oscillations that are locked to the phase of the drive. The fixed point solutions $(a_{1,0},a_{2,0},\Phi_0)$ of the three dynamical equations (\ref{amp1})-(\ref{phi}) correspond to periodic orbits of Eqs.~(\ref{amp_eq}) with frequency $\Omega_0/2$, where $\Omega_0=f_4(a_{1,0},a_{2,0},\Phi_0)$. The amplitudes and frequency of these solutions are shown in Fig.~\ref{fig2} as a function of the drive frequency for $g=10$, $\eta=1$, and a coupling $\Delta=7$, corresponding to a linear mode splitting of $\sqrt{1+2\Delta}-1\simeq2.87$ in units of the resonance frequency $\omega_0$.

We now demonstrate how the dynamics can be utilized to eliminate the phase noise of the driving oscillator.  We model this noise as white noise of intensity $F_{\Omega_{p}}$ in the frequency of the drive, replacing $\Omega_p$ with $\Omega_p+\Xi_p(T)$, and assuming that $\langle\Xi_p(T)\Xi_p(T')\rangle=F_{\Omega_{p}}\delta(T-T')$. This noise causes the phase of the driving source $\Phi_p$ to diffuse as $\langle[\Phi_p(T+\tau)-\Phi_p(T)]^2\rangle=F_{\Omega_{p}}|\tau|$. Additional noise in the amplitude of the drive signal can be suppressed by using a limiter or a highly saturated amplifier \cite{yurkePrl,yurkePra}. The output signal, which is obtained by mixing and filtering the response to this drive, is proportional to $\cos[(\omega_2-\omega_1)t+\Psi(T)]$. At the steady state of oscillation, the phase $\Psi$ can be expressed as $\Psi(T)=\Omega_0T+\psi(T)$, where $\psi$ is the small stochastic perturbation induced by the noise  $\Xi_p$. The phase perturbation $\psi$ can be calculated by solving the linearized version of the polar amplitude equations (\ref{polar}) spectrally, as was done for a single oscillator by Yurke \emph{et al.}~\cite{yurkePra}, or using the secular perturbation method of Demir \emph{et al.}~\cite{Demir00}.  This gives diffusion of the phase of the output signal, with the variance of the phase difference growing linearly in time \begin{eqnarray}\label{phaseDiff}
    &&\langle[ \psi(T+ \tau)-\psi(T)]^{2}\rangle=D_{\Omega_p}F_{\Omega_{p}}|\tau|,
\end{eqnarray}
where the diffusion constant $D_{\Omega_p}$ quantifies the sensitivity of the phase of the output signal to the noise~$\Xi_p$ in the drive frequency.  This expression applies in the limit that the time $\tau$ is much longer than the decay time onto the limit cycle for the noiseless oscillator. This approximation corresponds to frequency offsets close to the oscillator frequency. This is the relevant regime to explore since the width of the oscillator spectral peak is typically narrower than that of the driven resonator, whose width is determined by the quality factor, which in turn sets the decay time. The phase diffusion induced by white drive-frequency noise corresponds to a Lorenzian spectral peak of the output signal with a width of $D_{\Omega_p}F_{\Omega_{p}}$ times the resonator bandwidth $\epsilon\omega_0$.  Equation (\ref{phaseDiff}) is the expected result because after transients have decayed and the system has settled onto the limit cycle, the net effect of the stochastic dynamics is a Brownian motion of the free phase.

\begin{figure}[]
\begin{center}
   \includegraphics[width=1\columnwidth]{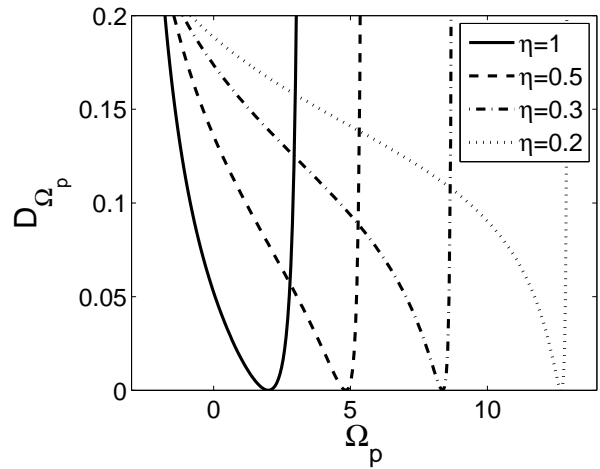}
   \end{center}
   \caption{\label{fig3} Phase diffusion of the output signal induced by frequency noise of unit strength for different values $\eta$ of the nonlinear damping. For each curve, $\Omega_p^{\textmd{max}}$ is the value of $\Omega_p$ for which the diffusion is zero. Note that $D_{\Omega_{p}}$ is also smaller than the phase diffusion of the driving oscillator along most of the curve, and that the solid curve is the squared derivative of the solid curve in Fig.~\ref{fig2}(b).}
\end{figure}

Without going into the details of the calculation, we note that the diffusion coefficient $D_{N}$ for a white noise source $N$ in Eqs.~(\ref{polar}) may be calculated as the squared scalar product of two vectors $D_{N}=(\textbf{v}_\bot\cdot\textbf{v}_{N})^2$.
The vector $\textbf{v}_\bot$ captures the phase sensitivity of the system, through the Jacobian matrix of Eqs.~(\ref{polar}) describing the linearized flow in the vicinity the limit-cycle. Specifically, it is the eigenvector of the transpose of this Jacobian matrix that corresponds to the zero eigenvalue. $\textbf{v}_{N}$ is the noise vector, whose $n$th entry is the term multiplying the noise source in the function $f_n$ ($\textbf{v}_{N}=(0,0,1,0)$ for the current example since $\Omega_p$ only appears in Eq.~(\ref{phi}) and it has a coefficient of magnitude one).
This description of the diffusion coefficient is a simplification of the general approach described in Ref.~\cite{Demir00} for systems in which motion along the limit-cycle is described by a steadily advancing phase variable $\Psi$ which does not affect the time evolution of the dynamical system, and thus the limit-cycle is represented by a fixed point in all other variables. In terms of the current example, this reducibility of the dynamical system is expressed by the fact that $\Psi$ does not appear on the right hand sides of Eqs.~(\ref{polar}). In the more general case the corresponding vectors $\textbf{v}_\bot(t)$ and $\textbf{v}_{N}(t)$ are time dependant: they are periodic, having the period of the limit-cycle, and the diffusion coefficient is given by the time average of their squared scalar product.

In systems such as the current one under consideration, calculating the zero mode of the transposed Jacobian matrix provides a way to calculate the phase diffusion that results from any white noise vector acting on a limit-cycle. However, if the noise originates from fluctuations in some parameter of the equations $p$ so that the noise vector is $\textbf{v}_{N}=\textbf{v}_p=\left(\frac{\partial f_1}{\partial p},\frac{\partial f_2}{\partial p},...,\frac{\partial f_n}{\partial p}\right)$, the long time phase diffusion is directly related to the dependence of the oscillation frequency on the parameter $p$ through
\begin{equation}\label{derivative}
    D_{p}=(\textbf{v}_\bot\cdot\textbf{v}_{p})^2=\left(\frac{d\Omega_0}{dp}\right)^2.
\end{equation}
The second equality follows from a perturbation analysis of the change in frequency $\delta\Omega_p$ due to a small perturbation $\delta p$ in the parameter $p$. Eq.~(\ref{derivative}) shows that to reduce the frequency stability degradation due to parameter noise we seek extremum points in the curves of the oscillation frequency \textit{vs.}~the noisy parameter. This possibility of complete noise elimination is due to the reducibility of the dynamical description, which makes the two vectors used for calculating the diffusion coefficient constant in time. In this case, it is possible to tune a single parameter and make these vectors orthogonal \footnote{For a general limit-cycle, the result equivalent to Eq.~(\ref{derivative}) separates into two equations: $D_{p}=t_p^{-1}\int_0^{t_p}[\textbf{v}_\bot(t)\cdot\textbf{v}_{p}(t)]^2dt$, and $t_p^{-1}\int_0^{t_p}\textbf{v}_\bot(t)\cdot\textbf{v}_{p}(t)dt=d\Omega_0/dp$, where $t_p$ is the period \cite{Demir00}. For this general case the two vectors need to be orthogonal at every point in time in order for the phase diffusion coefficient to be zero, and generally it is not possible to achieve this by tuning a single parameter.}.

We now apply the ideas of the previous paragraphs to our noise reduction setup, for which the noisy parameter is $p=\Omega_p$. Figure \ref{fig2}(b) shows an extremum we can exploit at the maximum of the frequency curve. By operating at this point, \textit{i.e.}\ at the drive frequency $\Omega_p^{\textmd{max}}$, the effect of drive frequency noise on the phase diffusion is eliminated, as shown in Fig.~\ref{fig3}, and the corresponding frequency spectrum is narrowed (to a sharp peak in the absence of other noise terms). It is also noteworthy that
for the parameters used to plot the diffusion coefficient $D_{\Omega_{p}}$ in Fig.~\ref{fig3}, the frequency stability is also improved along most of the curve, since $D_{\Omega_{p}}<1$, and thus the output frequency noise is less than input frequency noise.
An examination of the different curves in the figure indicates that larger nonlinear damping makes $\Omega_p^{\textmd{max}}$ more accessible experimentally, since this increases the frequency separation between the maximum $(d\Omega_0/d\Omega_p=0)$ and the saddle-node $(d\Omega_0/d\Omega_p=\infty)$. It is possible to control the nonlinear parameters of a resonator (both Duffing and damping terms) by adding a parametric feedback loop, as we have recently shown elsewhere \cite{pfo}.

\begin{figure}[]
\begin{center}
   \includegraphics[width=0.8\columnwidth]{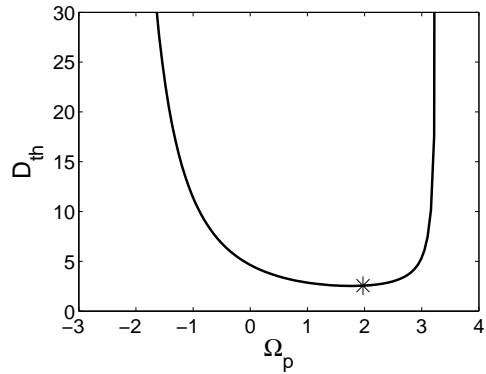}
   \end{center}
   \caption{\label{fig4} Thermal noise limit of the phase diffusion of the output signal, for $F_{\text{th}}=1$ and the same parameters as in Fig.~\ref{fig2}. The asterisk indicates the value at $\Omega_p^{\textmd{max}}$.}
\end{figure}

The ability of the device to clean phase noise is limited by the thermal noise floor. To discover this fundamental limit we add the complex thermal noise terms $\Xi_n=\Xi_{Rn}+i\Xi_{In}$ to each of Eqs.~(\ref{amp_eq}),
with the individual noise components white, uncorrelated, and of the same intensity $F_{\text{th}}$. Our analysis shows that the phase diffusion resulting from thermal noise is then given by Eq.~(\ref{phaseDiff}), but with the coefficients $D_{\Omega_p}F_{\Omega_{p}}$ replaced by $D_{\textmd{th}}F_{\text{th}}$, where
$D_{\textmd{th}}$ is plotted in Fig.~\ref{fig4}. In terms of the actual physical parameters, the lower phase diffusion limit is
$(k_{\textmd{B}}T/E_{c})D_{\text{th}}$ (in units of the bandwidth $\epsilon\omega_0$) with $k_\textmd{B}T$ the thermal energy, $E_{c}=\epsilon m^2\omega_0^4/\tilde{\alpha}$ the potential energy of the resonator element when driven to the Duffing critical amplitude, $\tilde{\alpha}$ the Duffing nonlinearity parameter (in units of force per volume), and $m$ the resonator mass (see Ref.\ \cite{LCreview}).
Since thermal noise does not originate from some fluctuating parameter, it cannot be eliminated by finding extremum points in the dependence of the output frequency. Nevertheless, it is possible to lower the thermal noise limit by reducing $D_{\textmd{th}}$. The idea is based on the fact that the direct effect of thermal noise on the phase variable is reduced at large amplitudes, so in the large amplitude limit, the only cause of phase diffusion is the conversion of noise from the amplitude to the phase (AM-PM conversion). For a single oscillator, this conversion is completely eliminated by operating at a point for which the resonator frequency is insensitive to the amplitude \cite{dykman90}, and this approach can be generalized to achieve amplitude-phase detachment in our system as well.

In conclusion, we have described a passive device that eliminates phase noise in oscillators. The device is made of two coupled resonators, driven parametrically in the non-degenerate mode by the output of a conventional oscillator. We find a driving frequency for which the resulting limit-cycle oscillation frequency is insensitive to the drive frequency and show that operating at this point eliminates the phase noise in the driving oscillator. We have discussed the operational limitation due to thermal noise, and have suggested ways to improve this limit. Along with the interesting physics it portrays, this device offers a practical way to handle the extensively studied, cardinal problem of oscillator phase noise.

This research was supported by DARPA through the DEFYS program.
\bibliography{nonDegenerateSubmission}

\end{document}